\documentclass[aps,prb,12pt,superscriptaddress
]{revtex4-2}

\usepackage{graphicx}
\usepackage{booktabs}
\usepackage{amsmath}
\usepackage{dcolumn}
\usepackage{url}
\usepackage{bm}
\usepackage{multirow}
\usepackage[hidelinks]{hyperref}

\usepackage{xcolor}

\definecolor{suppgreen}{RGB}{0,128,0}

\renewcommand{\thefigure}{\arabic{figure}}
\renewcommand{\thetable}{\arabic{table}}

\newcommand{\figref}[1]{%
  \textcolor{suppgreen}{Fig.~\ref{#1}}%
}

\makeatletter
\renewcommand{\fnum@figure}{%
  \textcolor{suppgreen}{\figurename~\thefigure}%
}
\renewcommand{\fnum@table}{%
  \textcolor{suppgreen}{\tablename~\thetable}%
}
\makeatother\makeatother
\makeatletter
\renewcommand\subsection{%
  \@startsection{subsection}{2}%
  {\parindent}%
  {1.5ex plus .3ex minus .2ex}%
  {-1em}%
  {\normalfont\normalsize\bfseries}%
}
\makeatother
\begin{document}
\title{Chiral Antiferromagnetism from Momentum-Space Resonance in a 2D Semiconductor}

\author{R.~Okuma}
\email{rokuma@issp.u-tokyo.ac.jp}
\affiliation{Okinawa Institute of Science and Technology Graduate University, Onna-son, Okinawa 904-0495, Japan}
\affiliation{Institute for Solid State Physics, University of Tokyo, Kashiwa, Chiba 277-8581, Japan}

\author{T.~Ikenobe}
\thanks{These authors contributed equally to this work.}
\affiliation{Institute for Solid State Physics, University of Tokyo, Kashiwa, Chiba 277-8581, Japan}

\author{Y.~Fujisawa}
\thanks{These authors contributed equally to this work.}
\affiliation{Okinawa Institute of Science and Technology Graduate University, Onna-son, Okinawa 904-0495, Japan}
\affiliation{Research Institute for Synchrotron Radiation Science (HiSOR), Hiroshima University, Higashi-Hiroshima, Hiroshima 739-0046, Japan}

\author{K.~Yamagami}
\affiliation{Okinawa Institute of Science and Technology Graduate University, Onna-son, Okinawa 904-0495, Japan}
\affiliation{Japan Synchrotron Radiation Research Institute, Sayo, Hyogo 679–5198, Japan}

\author{H. C. H.~Wu}
\affiliation{Clarendon Laboratory, University of Oxford, Parks Road, Oxford OX1 3PU, UK}

\author{T.~Nakamura}
\affiliation{Okinawa Institute of Science and Technology Graduate University, Onna-son, Okinawa 904-0495, Japan}

\author{Y.~Ihara}
\affiliation{Department of Physics, Faculty of Science, Hokkaido University, Sapporo 060-0810, Japan}

\author{H.~Ishikawa}
\affiliation{Institute for Solid State Physics, University of Tokyo, Kashiwa, Chiba 277-8581, Japan}
\affiliation{Department of Applied Chemistry, Faculty of Science Division 1, Tokyo University of Science, Shinjuku, Tokyo 162-8601, Japan}

\author{H.~Suwa}
\affiliation{Department of Physics, The University of Tokyo, Hongo, Tokyo 113-0033, Japan}

\author{H.~Ishizuka}
\affiliation{Department of Physics, Institute of Science Tokyo, Meguro, Tokyo 152-8551, Japan}

\author{Y.~Akagi}
\affiliation{Faculty of Core Research Natural Sciences Division, Ochanomizu University, Ohtsuka, Tokyo 112-8610, Japan}

\author{T.~Kaneko}
\affiliation{Department of Physics, The University of Osaka, Toyonaka, Osaka 560-0043, Japan}

\author{C.~H.~Hsu}
\affiliation{Okinawa Institute of Science and Technology Graduate University, Onna-son, Okinawa 904-0495, Japan}

\author{Y.~Obata}
\affiliation{Okinawa Institute of Science and Technology Graduate University, Onna-son, Okinawa 904-0495, Japan}

\author{N.~Tomoda}
\affiliation{Okinawa Institute of Science and Technology Graduate University, Onna-son, Okinawa 904-0495, Japan}

\author{M.~Dronova}
\affiliation{Okinawa Institute of Science and Technology Graduate University, Onna-son, Okinawa 904-0495, Japan}

\author{K.~Nagasawa}
\affiliation{Institute for Solid State Physics, University of Tokyo, Kashiwa, Chiba 277-8581, Japan}

\author{H.~Saito}
\affiliation{Institute for Solid State Physics, University of Tokyo, Kashiwa, Chiba 277-8581, Japan}

\author{D.~Ueta}
\affiliation{Institute of Materials Structure Science, High Energy Accelerator Research Organization (KEK), Tsukuba, Ibaraki 305-0801, Japan}

\author{H.~Sagayama}
\affiliation{Institute of Materials Structure Science, High Energy Accelerator Research Organization, Tsukuba, Ibaraki 300-3256, Japan}

\author{J.~Yamaura}
\affiliation{Institute for Solid State Physics, University of Tokyo, Kashiwa, Chiba 277-8581, Japan}

\author{M. ~Arita}
\affiliation{Research Institute for Synchrotron Radiation Science (HiSOR), Hiroshima University, Higashi-Hiroshima, Hiroshima 739-0046, Japan}

\author{K. ~Yogendra}
\affiliation{Research Institute for Synchrotron Radiation Science (HiSOR), Hiroshima University, Higashi-Hiroshima, Hiroshima 739-0046, Japan}

\author{S. ~Ideta}
\affiliation{Research Institute for Synchrotron Radiation Science (HiSOR), Hiroshima University, Higashi-Hiroshima, Hiroshima 739-0046, Japan}

\author{K.~Kindo}
\affiliation{Institute for Solid State Physics, University of Tokyo, Kashiwa, Chiba 277-8581, Japan}

\author{T.~Nakajima}
\affiliation{Institute for Solid State Physics, University of Tokyo, Kashiwa, Chiba 277-8581, Japan}

\author{S.~J.~Blundell}
\affiliation{Clarendon Laboratory, University of Oxford, Parks Road, Oxford OX1 3PU, UK}

\author{T.~Kondo}
\affiliation{Institute for Solid State Physics, University of Tokyo, Kashiwa, Chiba 277-8581, Japan}

\author{K. ~Shimada}
\affiliation{Research Institute for Synchrotron Radiation Science (HiSOR), Hiroshima University, Higashi-Hiroshima, Hiroshima 739-0046, Japan}
\affiliation{WPI-SKCM2, Hiroshima University, Higashi-Hiroshima, Hiroshima 739-8530, Japan.}
\affiliation{Research Institute for Semiconductor Engineering, Hiroshima University, Higashi-Hiroshima, Hiroshima 739-8530, Japan.}

\author{Y.~Okamoto}
\affiliation{Institute for Solid State Physics, University of Tokyo, Kashiwa, Chiba 277-8581, Japan}

\author{Y.~Okada}
\email{yoshinori.okada@oist.jp}
\affiliation{Okinawa Institute of Science and Technology Graduate University, Onna-son, Okinawa 904-0495, Japan}

\begin{abstract}
Understanding the principles governing the emergence of chiral quantum phases is a fundamental challenge, not only for uncovering new mechanisms of quantum-state formation but also for realizing giant electronic responses and transport phenomena arising from chirality and topology. While Fermi-surface instabilities in metals can stabilize complex ordered states through multiple competing scattering channels, their microscopic origin is often obscured by the complexity of the underlying electronic structure, limiting the development of general microscopic design principles. Here, we introduce a complementary strategy based on the simplicity of semiconductor band extrema. Using the layered van der Waals semiconductor GdGaI, whose low-energy electronic structure consists of simple electron and hole valleys, we discover the spontaneous emergence of an intertwined chiral triple-$q$ antiferromagnetic state accompanied by a cooperative reconstruction of the electron-hole band edges, beyond the conventional expectation of a single-$q$ ground state. This collective reconstruction generates substantial momentum-space Berry curvature, giving rise to a pronounced spontaneous anomalous Hall effect despite the semiconducting character and negligible net magnetization. Remarkably, this chiral state is realized within an atomically well-defined ($\approx2a$), topologically nontrivial magnetic texture, showing that such collective quantum states can emerge at an exceptionally small length scale from a simple two-dimensional magnetic semiconductor. More broadly, our results introduce a remarkably simple design concept for chiral quantum matter: using simple semiconductor band extrema as building blocks for resonance-like interplay in momentum space, providing a route to Berry curvature, topological transport, and emergent quantum phases.
\end{abstract}
\maketitle
\clearpage
\section*{Introduction}
Multi-$q$ ordered states, in which multiple symmetry-related ordering vectors coexist, have attracted growing interest because they can host emergent phenomena beyond conventional single-$q$ phases such as topological real-space textures\cite{roessler2006spontaneous,muhlbauer2009skyrmion,yu2010real,heinze2011spontaneous,seki2012observation,kezsmarki2015neel,yu2018transformation,fujishiro2019topological,kurumaji2019skyrmion,hirschberger2019skyrmion,khanh2020nanometric,zheng2023hopfion} and unconventional electromagnetic responses associated with Berry curvature\cite{ohgushi2000spin,taguchi2001spin,machida2010time,neubauer2009topological,tokura2021magnetic}. A powerful paradigm for realizing such states is the amplification of electronic instabilities through Fermi-surface scattering\cite{martin2008itinerant,akagi2010spin,Akagi_2012,hayami2014multiple,jiang2015chiral,wang2020skyrmion,bouaziz2022fermi,paddison2022magnetic,dong2025pseudogap,arai2026origin}. The rich electronic structures of kagom\'e metals and other correlated materials have revealed an extraordinary diversity of intertwined and multi-$q$ phases arising from this mechanism\cite{ortiz2019new,Jiang_2021,Mielke_2022,Kang_2022,Denner_2021,Park_2021}. Yet this richness also presents a fundamental challenge since multiple nesting vectors, scattering channels, and competing instabilities often coexist near the Fermi level, making it difficult to disentangle the primary instability from its secondary consequences and to identify the minimal ingredients required for spontaneous chiral symmetry breaking. This naturally raises a fundamental question: {\it can spontaneous chiral symmetry breaking emerge from a much simpler electronic structure}?

Narrow-gap semiconductors provide an attractive setting to address this question because their low-energy electronic structure is naturally simplified. Near the band edge, only a small number of electron and hole valleys participate in the instability, sharply restricting the available ordering wave vectors and scattering channels\cite{Kohn_1967,Jerome_1967,HALPERIN_1968,Cercellier_2007,Monney_2009,Kogar_2017,Wakisaka_2009,Lu_2017,Sun_2021}. Once a single ordering channel develops and lowers the electronic energy, there is no obvious reason for the remaining symmetry-related channels to condense simultaneously (\figref{Fig1}a). This intuitive expectation, however, leaves open an important possibility. If multiple symmetry-related instabilities cooperate, a multi-$q$ state may become energetically favorable despite the apparent preference for single-$q$ order (\figref{Fig1}b). The situation becomes even richer when magnetism is involved, because magnetic and electronic instabilities need not act independently but may instead reinforce one another in selecting the symmetry of the ordered state\cite{hayami2014multiple,bouaziz2022fermi,Park_2021,Vylet_2026,Guzman_2026}. This suggests a different perspective on spontaneous symmetry breaking, in which the ordered state is determined not solely by the strongest instability but by the cooperative interplay among multiple symmetry-related ordering channels. Whether such momentum-space cooperation can spontaneously generate a chiral multi-$q$ state in a semiconductor with an exceptionally simple electronic structure remains an open question (\figref{Fig1}b–c).

Here we address this question using the layered antiferromagnetic semiconductor GdGaI (GGI) (\figref{Fig1}d)\cite{Okuma_2021,Posey_2024,Lukachuk_2007,okuma2024,Kaneko_2026}, whose simple low-energy electronic structure provides an ideal platform for isolating the essential instability. Density-functional-theory calculations show that states near the Fermi level are derived primarily from Ga-$4p$ and Gd-$5d$ orbitals and form an indirect band gap between symmetry-related extrema (\figref{Fig1}e). Conventional considerations predict that such a system should adopt a single-$q$ ordered state: reconstructing a single pair of band extrema already provides the dominant electronic energy gain, while conventional bilinear magnetic interactions on the triangular lattice tend to favor single-$q$ spiral or collinear orders\cite{RASTELLI19791,jolicoeur1990ground,chubukov1992order}. Whether such a minimal electronic system can nevertheless self-organize into a more complex chiral multi-$q$ state has remained unexplored.

Using the first high-quality single crystals of GdGaI (\figref{Fig1}f), we uncover a strikingly different scenario. Rather than remaining in a single-$q$ state, GdGaI develops an intertwined chiral triple-$q$ phase in which magnetic order, electronic reconstruction, and charge modulation emerge cooperatively. The three ordering channels form a closed scattering network connecting the semiconductor band extrema, thereby providing an energetic distinction from single-$q$ reconstruction and acquiring chirality through its magnetic component. The resulting electronic state develops finite Berry curvature and a large spontaneous anomalous Hall response despite its low carrier density. In this study, we propose this cooperative closed-loop phenomenon as chiral resonance, demonstrating that complex chiral order can emerge from an exceptionally simple semiconductor band structure.

\section*{Results and Discussion}
\subsection*{Spontaneous Hall response and electronic reconstruction.}
A first indication comes from a comparison of the electronic structures at 300 and 14 K. At 300 K, Angle resolved photoemission spectroscopy (ARPES) reveals a simple semiconductor band structure with the conduction-band minima located at the $M$ points (\figref{Fig1}g–i). Upon cooling to 14 K, however, the electronic structure is substantially reconstructed: a hole-like band appears at $M$, with its maximum nearly degenerate with that at $\Gamma$. The emergence of this folded band indicates a characteristic reconstruction wave vector $|Q| = |a^*/2|$. This striking contrast between the high- and low-temperature electronic structures raises an immediate question: when and how does this reconstruction develop upon cooling?

Transport and magnetization provide an important clue as to where to look. Weak ferromagnetic order develops below $T_\mathrm{WF}\approx$ 28 K (\figref{Fig1} j), with only a small out-of-plane moment (\figref{Fig1}k), accompanied by an anomaly in the resistivity (\figref{Fig1}l). More remarkably, a large spontaneous Hall response emerges in the same low-temperature regime. The Hall resistivity $\rho_{yx}$ exhibits pronounced hysteresis and a step-like sign reversal around zero field (\figref{Fig1}m), establishing a spontaneous anomalous Hall effect (See Supplementary Note 4 for transport analysis). The zero-field Hall resistivity reaches $\pm0.5~\mathrm{m}\Omega\cdot$cm, comparable to or even exceeding values reported in anomalous Hall antiferromagnets\cite{Nakatsuji_2015,Kiyohara_2016,Nayak_2016,fujishiro2021giant,takagi2023spontaneous,park2023tetrahedral,LeeBroken2024,TakagiFeS_2024,Yamadapwave_2025}. The substantial spontaneous Hall response despite the tiny net magnetization ($<0.02\mu_\mathrm{B}$) points to an unconventional chiral electronic state.

These observations motivate us to ask whether the dramatic electronic reconstruction, magnetic ordering, and spontaneous Hall response share a common origin. We therefore first determine the underlying magnetic structure and then track the electronic reconstruction in detail as a function of temperature across the magnetic transition.

\subsection*{Triple-$q$ magnetic/charge coupling.}
Neutron diffraction, $\mu$SR, and nuclear magnetic resonance (NMR) jointly establish the magnetic ground state below $T_\mathrm{WF}$ with each probe providing complementary information on the magnetic order. Comparing neutron diffraction patterns measured at 5 K (\figref{Fig2}a) and 52 K (\figref{Fig2}b), we observe prominent magnetic Bragg peaks at $q$ = (-1/2, 0, 0) and (0, -1/2, 0) below $T_\mathrm{WF}$, corresponding to a doubling of the crystallographic unit cell. Their intensities at 5 K substantially exceed those of the nuclear Bragg peaks, demonstrating robust magnetic order with characteristic wave vectors connecting the electron and hole band extrema in the Brillouin zone (\figref{Fig1}c). Because neutron diffraction alone cannot unambiguously resolve changes between magnetic structures with the same $M$-point propagation vectors, we complemented it with $\mu$SR measurements, which probe the local distribution of the internal magnetic fields. The $\mu$SR asymmetry exhibits a well-defined oscillation exclusively below $T_\mathrm{WF}$ (\figref{Fig2}c), confirming the onset of static long-range magnetic order (see Supplementary Note 5 for the magnetic correlations above $T_\mathrm{WF}$). Finally, independent symmetry analysis combined with NMR uniquely identifies the ordered state as the $P\bar{3}m'1$ phase, consisting of a coherent superposition of three symmetry-related ordering vectors (\figref{Fig2}d and Supplementary Note 5). The internal field extracted from zero-field $\mu$SR at 5 K ($\approx$200 MHz) is consistent with this magnetic structure.

Let $S(Q_i)$ denote the three magnetic components at the symmetry-related $M$ points $Q_i$. Because $Q_1 + Q_2 + Q_3 = 0$ modulo a reciprocal-lattice vector, their coherent triple-$q$ condensation permits uniform composite order as well as spin–charge couplings at the same $M$-point wavevectors\cite{Barros_2014}. Translational and time-reversal symmetries allow a third-order free-energy term of the form $\rho(Q_1)[S(Q_2)\cdot S(Q_3)]$, together with its cyclic permutations, where $\rho(Q_1)$ denotes the charge modulation\cite{HayamiCDW2021,Park_2021}. The resulting charge modulation belongs to the same $M$-point star and has the same $2\times2$ periodicity as the magnetic order, with its phase locked to the coherent triple-$q$ state. Consistently, STM observes a clear $2\times2$ electronic modulation (\figref{Fig2}e), whereas the absence of corresponding synchrotron X-ray superlattice reflections (Supplementary Note 2) and recent Raman spectroscopy experiments constrain any accompanying significant lattice distortion\cite{Jiang_2026}.

The coherent noncoplanar triple-$q$ state also supports a time-reversal-odd $q = 0$ composite, which can couple to the symmetry-allowed $c$-axis magnetization of the $P\bar{3}m'1$ state. Because the dominant magnetic order resides at the three symmetry-related $M$ points, we regard the triple-$q$ antiferromagnetic order as primary and the small net moment as a concomitant, or parasitic, ferromagnetic component of the predominantly antiferromagnetic state. Such a small net moment is also observed in the Co$_{1/3}$(Nb,Ta)S$_2$, featuring the same in-plane spin arrangement\cite{takagi2023spontaneous}. This symmetry framework connects the magnetic and electronic reconstructions; whether they share a common onset is examined below through their temperature evolution.

\subsection*{Triple-$q$ band reconstruction.}
We used temperature-dependent synchrotron ARPES to test whether the semiconductor band edge reconstructs concomitantly with the onset of triple-$q$ order. Because the valence-band maximum is located at the $\Gamma$ point (\figref{Fig1}i), this momentum provides the most sensitive probe of the low-energy electronic structure governing the semiconducting state. We therefore carried out high-resolution measurements focusing on this region. \figref{Fig3}a,b present the electronic structure at 300 K and 9 K, respectively, together with three-dimensional visualizations of the band dispersion (\figref{Fig3}c,d), which provide a more intuitive view of the band evolution. To quantitatively track the reconstruction, we further measured the temperature dependence of the ARPES dispersion along the high-symmetry $\Gamma–M$ direction (\figref{Fig3}e), extracted the band energies, and summarized their evolution in \figref{Fig3}f. Here, band energy is extracted from a symmetrized energy distribution curve, which basically corresponds to peak energy on the single particle spectral function. Temperature-dependent ARPES reveals a pronounced reconstruction of the valence band upon cooling. The most striking signature is a change in the band geometry: the valence-band maximum evolves into a characteristic camelback (M-shaped) dispersion, with minima located at finite momenta $k_H \approx \pm$0.13 \AA$^{-1}$ (\figref{Fig3}f).

We assess the relationship between the band reconstruction and the magnetic transition using two characteristic energy scales: $|E_\Gamma(T)|$, describing the overall evolution of the semiconductor band edge, and $|E_\Gamma(T) – E_{k_H}(T)|$ quantifying the camelback deformation (\figref{Fig3}g). Whereas $|E_\Gamma(T)|$ evolves gradually upon cooling (\figref{Fig3}h), the camelback deformation exhibits a pronounced onset at $T_\mathrm{WF}$ (\figref{Fig3}i). Comparison with the magnetic order parameter (\figref{Fig3}j) demonstrates that the camelback reconstruction develops simultaneously with the triple-$q$ magnetic order, establishing a direct coupling between the semiconductor band edge and magnetism. The same temperature evolution is reproducibly observed using independent light sources on differently cleaved surfaces in multiple ARPES measurements, confirming that the reconstruction is an intrinsic bulk property of GdGaI. Although the precise onset temperatures exhibit small probe/sample-dependent variations, these are secondary to the central conclusion in this study. 

\subsection*{Numerical calculation on the triple-$q$ intertwined magnetic state.}
A minimal Kondo-lattice model reproduces both the magnetic phase sequence and the accompanying electronic reconstruction. The model describes itinerant band-edge carriers exchange-coupled to localized $4f$ moments on a triangular bilayer, with the localized moments interacting through bilinear Heisenberg exchange (\figref{Fig4}a). At the classical-energy level of the bilinear spin Hamiltonian, the single-$q$ and triple-$q$ states are degenerate. This degeneracy is lifted by thermal fluctuations and by coupling to the itinerant carriers. Semiclassical finite-temperature simulations yield, upon cooling, an intermediate single-$q$ phase followed by a coherent triple-$q$ low-temperature phase. Concomitantly, the three electron pockets at the $M$ points are folded to the Brillouin-zone center and hybridize with the hole band, reproducing the key features of the experimentally observed electronic reconstruction (\figref{Fig4}b,c; see Supplementary Note 8 for details).

The experimentally determined noncoplanar triple-$q$ structure also provides a microscopic origin for the anomalous Hall response. We evaluate its Berry curvature using a complementary model that exchange-couples this magnetic structure to the itinerant electrons (Supplementary Note 8). Remarkably, the reconstructed bands develop finite Berry curvature even without spin-orbit coupling and in the absence of a net magnetization (\figref{Fig4}d). The calculation shows that the chiral triple-$q$ reconstruction provides a microscopic route to the spontaneous anomalous Hall effect. Experimentally, the anomalous Hall conductivity follows an approximate scaling relation of $|\sigma_{xy}|\propto \sigma_{xx}^{1.6}$, extending the established anomalous Hall scaling into the semiconducting regime (\figref{Fig4}e)\cite{Miyasato_2007,Tian_2009,OnodaAHE2006}. Despite its low carrier density ($<10^{19}~\mathrm{cm}^{-3}$), samples with sub-metallic conductivity reach $|\sigma_{xy}|\approx 1~\Omega^{-1}\cdot$cm$^{-1}$, and Hall angle of 2\%, comparable to those of metallic anomalous Hall antiferromagnets. Thus, the chiral electronic response remains substantial despite the low carrier density and semiconducting transport.

\subsection*{Chiral resonance in momentum space.}
The stabilization of triple-$q$ order in GGI may arise cooperatively from its charge and spin degrees of freedom. Unlike conventional itinerant multiple-$q$ magnets, GGI is a low-carrier semiconductor, suggesting a distinct band-edge mechanism. Because $Q_1 + Q_2 + Q_3 = 0$, the three ordering vectors form a closed scattering path in momentum space, permitting a cubic invariant involving all three reconstruction amplitudes\cite{Park_2021}. For the $\Gamma + 3M$ reconstruction relevant to GGI, a minimal scalar-potential model gives a third-order term proportional to the product of the three charge-scattering amplitudes (Supplementary Note 8). This term is absent in a single-$q$ state and can provide an additional electronic stabilization channel for coherent triple-$q$ order (\figref{Fig1}a-b). The three-component modulation observed by STM suggests a cooperative spin–charge reconstruction in which coherence between the three ordering channels enables electronic scattering processes unavailable to any single component.
This cooperative spin–charge state is reminiscent of resonance stabilization in aromatic compounds, where coherent electron delocalization over a closed bonding network lowers the electronic energy\cite{Huckel_1931,Pauling_1934,Mo_2009}. In GGI, however, the corresponding network is formed not by chemical bonds in real space but by successive scattering among electronic states in momentum space, giving rise to what we term momentum-space resonance. The same coherent triple-$q$ configuration produces a noncoplanar magnetic structure with finite scalar spin chirality. Although the electronic resonance energy and the spin chirality are distinct physical quantities, both originate from the simultaneous, phase-coherent presence of the three symmetry-related $q$ components. We refer to this combination of momentum-space electronic resonance and chiral magnetic order as chiral resonance (\figref{Fig4}f). Because the relevant electronic states are concentrated near a small number of simple semiconductor band extrema, GdGaI provides a particularly transparent realization of this physics, showing that cooperative chiral quantum matter can emerge without a complex metallic Fermi surface.

\section*{Summary}
In summary, we identify chiral resonance in the van der Waals antiferromagnetic semiconductor GdGaI, where magnetic order, charge modulation, and the semiconductor band edge reconstruct cooperatively into a chiral triple-$q$ state. This unusual form of multi-$q$ order produces finite Berry curvature and a spontaneous anomalous Hall response despite the low-carrier-density semiconducting state.
\section*{Acknowledgments}
R.O. and Y.O. acknowledge useful discussions with Y. Motome, C. D. Batista, Y. Niimi, N. Jiang, T. Higashihara, Y. Ohta and M. Ochi. R.O. acknowledges R. Coldea and D. Antoniou for their support for single crystal X-ray diffraction experiments. Simulations were performed using computational resources of the Supercomputer Center at the Institute for Solid State Physics, the FUJITSU Supercomputer PRIMEHPC FX1000 and FUJITSU Server PRIMERGY GX2570 (Wisteria/BDEC-01) at the Information Technology Center, the University of Tokyo. This work was supported by JSPS KAKENHI (Grant Nos. 22K03508, 23K19027, 24H00191, 24H01609, 24K00546, 24K06939, 25K17337, and 26K17091), JST (Grant No. JPMJPF2221), JST CREST (Grant No. JPMJCR24I1), JST PRESTO (Grant Nos. JPMJPR2251, JPMJPR2591, JPMJPR2593), JST ASPIRE (Grant No. JPMJAP2314), and the Center of Innovation for Sustainable Quantum AI (SQAI). Crystal structures were made using VESTA\cite{momma2011vesta}.
\section*{Figures}
\begin{center}
    \includegraphics[width=\columnwidth]{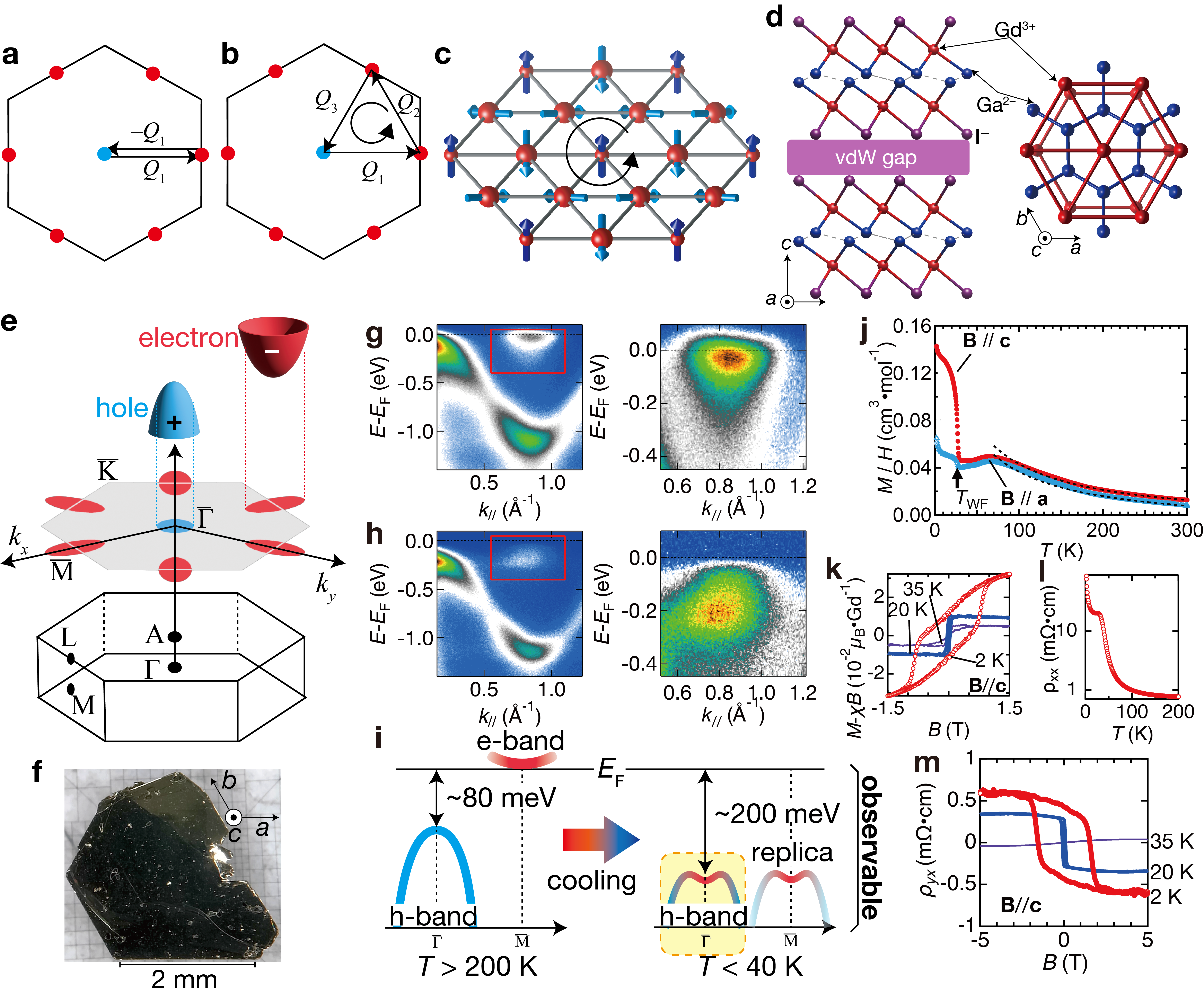}
\end{center}

\refstepcounter{figure}
\label{Fig1}
\noindent
\textbf{\textbf{FIG.~\thefigure.} Chiral triple-$q$ semiconductor in GGI.}
{\bf a,} Schematic electronic reconstruction induced by a single-$q$ instability in a semiconductor. The ordering hybridizes electronic states connected by a single ordering vector, leading to conventional pairwise gap formation near the band edge.
{\bf b,} Momentum-space structure of hybridization in the triple-$q$ case. The three ordering vectors cooperatively connect electronic states through the closed path $\Gamma\rightarrow M_i\rightarrow M_j\rightarrow \Gamma$ producing a loop-like hybridization process absent in the single-$q$ case.
{\bf c,} The real-space spin texture, where the superposition of three spin modulations on the triangular lattice forms a noncoplanar magnetic structure with finite scalar spin chirality.
{\bf d,} Crystal structure of GGI showing Gd (red), Ga (blue), and I (purple) atoms. Gd carries a localized magnetic moment corresponding to spin-7/2. The local structure highlights a Ga honeycomb layer sandwiched between triangular Gd layers (right).
{\bf e,} Schematic electronic structure of GGI. Hole and electron bands are separated by a finite indirect gap, with the valence band maximum at $\Gamma$ and the conduction band minima at the $M$ point.
{\bf f,} Optical microscope image of a single crystal of GGI with crystallographic axes indicated.
{\bf g,h,} ARPES spectra along high symmetry line $\Gamma-M$ at 300 K (g) and 14 K (h). The zoomed-in ARPES images around the $M$ point (highlighted in the red square) are shown in the right panel.
{\bf i,} The schematic band structure, compared at 300 K and 14 K.
{\bf j,} Magnetic susceptibility as a function of temperature measured along the in-plane and out-of-plane directions under an applied field of 1 T. The black line shows a Curie–Weiss fit at high temperatures with a Weiss temperature of -35 K. Below $T_\mathrm{WF} \sim$ 28 K, long-range magnetic order develops with a small spontaneous moment characteristic of antiferromagnetism from localized 4$f$ electrons.
{\bf k,} Magnetization processes across $T_{\mathrm{WF}}$. A linear-in-field contribution is subtracted from the magnetization to highlight the weak-ferromagnetism.
{\bf l,} Temperature dependence of the in-plane resistivity $\rho_{xx}$. Below $T_\mathrm{WF}$ a downward deviation from the overall insulating behavior appears.
{\bf m,} Field dependence of the Hall resistivity $\rho_{yx}$ measured at 2 K, 20 K, and 35 K. A spontaneous anomalous Hall effect appears only below $T_\mathrm{WF}$.
\clearpage
\begin{center}
    \includegraphics[width=10cm]{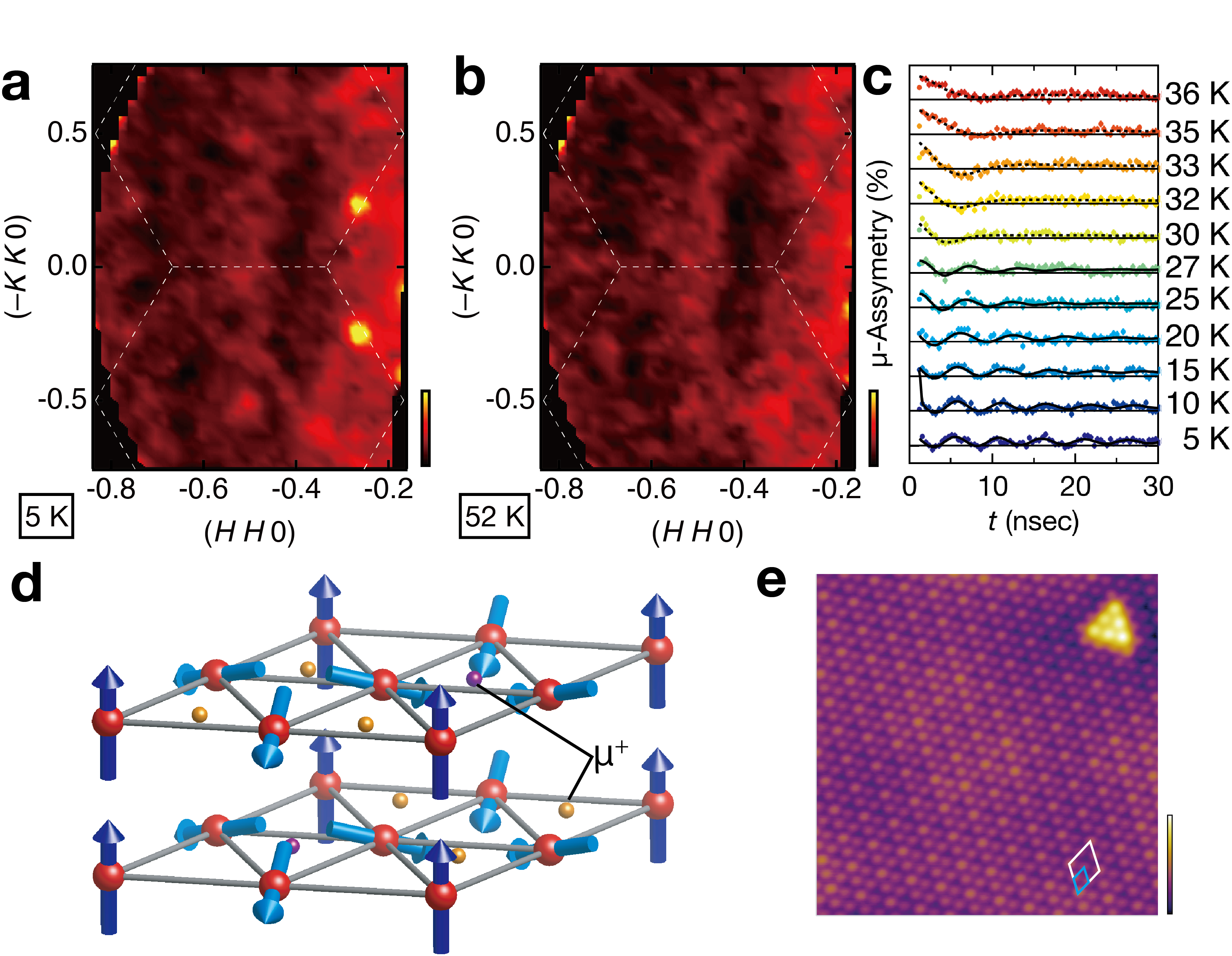}
\end{center}

\refstepcounter{figure}
\label{Fig2}
\noindent
\textbf{\textbf{FIG.~\thefigure.} Triple-$q$ magnetic and charge ordering in GGI. 
}
{\bf a,b} Neutron diffraction on a single crystal of GGI. Several magnetic Bragg peaks at (–1/2, 0, 0) and (0, –1/2, 0) appear at 5~K (a) but disappear at 52~K (b). 
{\bf c,} Zero-field $\mu$SR asymmetry showing a spontaneous oscillation below $T_\mathrm{WF}$. 
{\bf d,} Proposed triple-$q$ magnetic structure consistent with neutron diffraction, zero-field $\mu$SR, and NMR measurements. Red spheres and blue/skyblue arrows represent Gd atoms and magnetic moments, respectively. The muon stopping sites, which are close to the center of mass of Gd triangles, are shown in orange and purple spheres. Mostly the orange muon feels the large internal field from Gd. 
{\bf e,} The STM topographic image acquired at 4 K (set point: 400 pA, sample bias: +1 V). The characteristic contrast observed in the upper right corner (likely due to defects or impurities beneath the topmost iodine layer) confirms the isotropic tip shape, ensuring that the overall topographic image accurately reflects intrinsic GGI’s features. The sky-blue and white rhombi correspond to the unit cell of the original and $2\times2$ unit cell, respectively.

\clearpage

\begin{center}
    \includegraphics[width=\columnwidth]{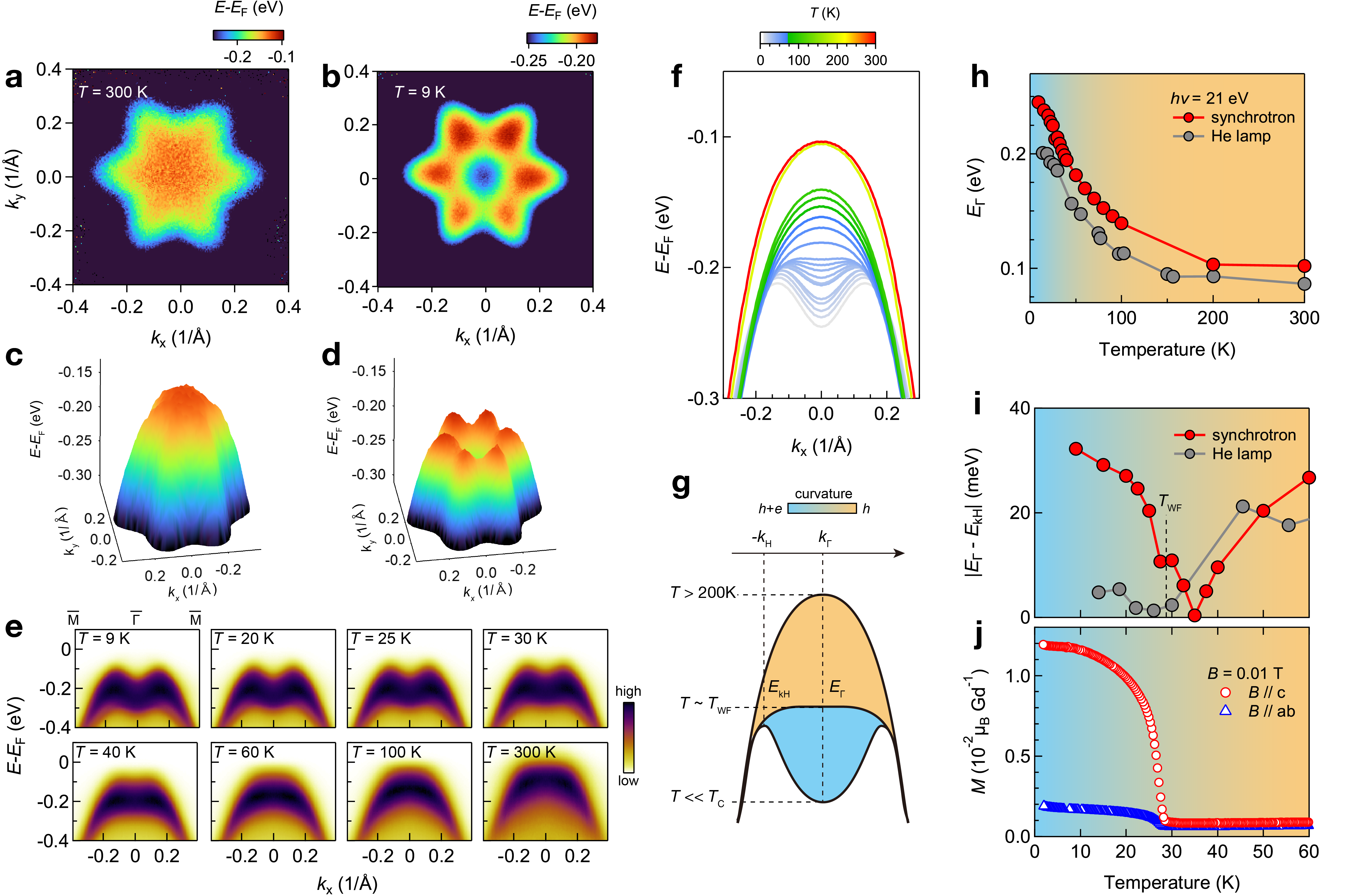}
\end{center}

\refstepcounter{figure}
\label{Fig3}
\noindent
\textbf{\textbf{FIG.~\thefigure} Emergence of camel-back structure on the hole-band dispersion. 
}
{\bf a,b}Momentum-space maps of the fitted hole-band peak energy at 300~K (a) and 9~K (b). At each momentum, the peak energy was determined by fitting the energy distribution curve with a Gaussian function.
{\bf c,d,} The corresponding band structure visualized in 3D image at 300~K (c) and 9~K (d).
{\bf e,} Temperature-dependent ARPES spectra along the $M-\Gamma-M$ direction. Temperatures are indicated in each panel.
{\bf f,} Band dispersions extracted from e, coloured according to temperature.
{\bf g,} Schematic of the temperature-dependent evolution of the hole-band curvature. $E_\Gamma$ and $E_{k_H}$ denote the fitted peak energies at $k$ = 0 and –0.13 \AA$^{-1}$, respectively.
{\bf h,i,} Temperature dependence of $E_\Gamma$ (h) and $|E_\Gamma-E_{k_H}|$ (i), obtained from two independent data sets acquired using synchrotron radiation and a helium discharge lamp.
{\bf j,} Temperature-dependent magnetization measured under a magnetic field of 0.01 T applied parallel to the c axis and the ab plane. The dashed line marks $T_\mathrm{WF}$. The background shading in h–j represents the evolution of the hole-band curvature illustrated in g.
\clearpage

\begin{center}
    \includegraphics[width=\columnwidth]{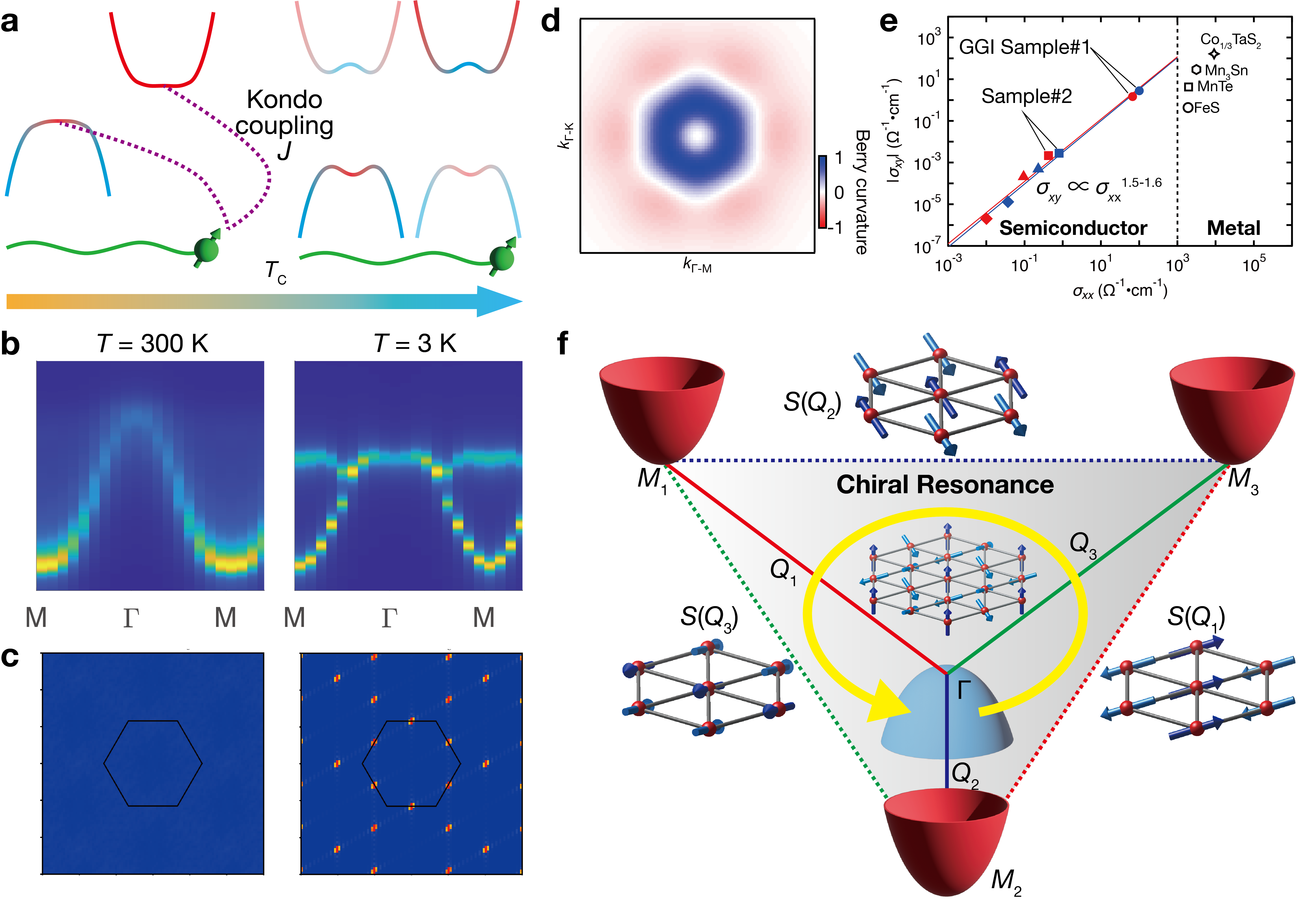}
\end{center}

\refstepcounter{figure}
\label{Fig4}
\noindent
\textbf{\textbf{FIG.~\thefigure} Microscopic origin of chiral resonance in GdGaI.
}
{\bf a,} Schematic illustration of band evolution induced by Kondo coupling $J$, which describes a spin interaction between localized moments and conduction/valence electrons. At low temperatures, this interaction leads to hybridization that reconstructs the electronic structure. The Kondo coupling effectively mediates hybridization between electron- and hole-like states near the Fermi level.
{\bf b,} Simulated spectral function showing the temperature evolution at 300~K (left) and 3~K (right), highlighting the emergence of hybridized bands. The triple-$q$ magnetic ordering temperature is 30 K. The discrete appearance of the spectra along the momentum originates from finite-size effects in the numerical calculation. 
{\bf c,} Spin structure factor of the localized moments $S(k)$, calculated using the Kondo-lattice model at 300 K (left) and 3~K (right).
{\bf d,} Momentum-space distribution of Berry curvature for a triple-$q$ spin configuration. The band reconstruction near the Fermi level generates finite Berry curvature, providing the microscopic origin of the anomalous Hall effect. 
{\bf e,} Comparison of anomalous Hall conductivity $|\sigma_{xy}|$ as a function of longitudinal conductivity $\sigma_{xx}$ for various compounds. Filled red and blue symbols represent data for GGI measured at 5 K and 15 K, respectively; different symbol shapes correspond to different samples. Solid lines are fits to $|\sigma_{xy}|\propto\sigma_{xx}^{1.6}$ at each temperature. Open symbols denote literature data for Co$_{1/3}$TaS$_2$ (2 K)\cite{takagi2023spontaneous}, Mn$_3$Sn (300 K)\cite{Nakatsuji_2015}, MnTe (2 K)\cite{LeeBroken2024}, and FeS (300 K)\cite{TakagiFeS_2024}.
{\bf f,} Schematic of the proposed chiral resonance in the triple-$q$ state. Each spin component $S(Q_i)$ scatters electronic states both between $\Gamma$ and $M_i$ (solid lines) and between the other two $M$ valleys (dotted lines), with the different line styles representing the inequivalent scattering amplitudes of the $\Gamma-M_i$ and $M_i–M_j$ channels. The coherent coexistence of the three $Q_i$ components forms a noncoplanar triple-$q$ texture with finite scalar spin chirality and a closed network of momentum-space scattering paths. Interference around these loops imparts a geometric phase to the carriers, generating Berry curvature.
\clearpage
\section*{References}
\bibliography{GGI_main_formatted}%

@article{roessler2006spontaneous,
  author    = {U. K. Roessler and A. N. Bogdanov and C. Pfleiderer},
  title     = {{Spontaneous Skyrmion Ground States in Magnetic Metals}},
  journal   = {Nature},
  volume    = {442},
  number    = {7104},
  pages     = {797--801},
  year      = {2006},
  publisher = {Nature Publishing Group UK London},
}

@article{muhlbauer2009skyrmion,
  author    = {S. Muhlbauer and B. Binz and F. Jonietz and C. Pfleiderer and A. Rosch and A. Neubauer and R. Georgii and P. Boni},
  title     = {{Skyrmion Lattice in a Chiral Magnet}},
  journal   = {Science},
  volume    = {323},
  number    = {5916},
  pages     = {915--919},
  year      = {2009},
  publisher = {American Association for the Advancement of Science},
}

@article{yu2010real,
  author    = {X. Z. Yu and Y. Onose and N. Kanazawa and J. H. Park and J. H. Han and Y. Matsui and N. Nagaosa and Y. Tokura},
  title     = {{Real-Space Observation of a Two-Dimensional Skyrmion Crystal}},
  journal   = {Nature},
  volume    = {465},
  number    = {7300},
  pages     = {901--904},
  year      = {2010},
  publisher = {Nature Publishing Group UK London},
}

@article{heinze2011spontaneous,
  author    = {S. Heinze and K. von Bergmann and M. Menzel and J. Brede and A. Kubetzka and R. Wiesendanger and G. Bihlmayer and S. Bl{\"u}gel},
  title     = {{Spontaneous Atomic-Scale Magnetic Skyrmion Lattice in Two Dimensions}},
  journal   = {Nat. Phys.},
  volume    = {7},
  number    = {9},
  pages     = {713--718},
  year      = {2011},
  publisher = {Nature Publishing Group UK London},
}

@article{seki2012observation,
  author    = {S. Seki and X. Z. Yu and S. Ishiwata and Y. Tokura},
  title     = {{Observation of Skyrmions in a Multiferroic Material}},
  journal   = {Science},
  volume    = {336},
  number    = {6078},
  pages     = {198--201},
  year      = {2012},
  publisher = {American Association for the Advancement of Science},
}

@article{kezsmarki2015neel,
  author    = {I. K{\'e}zsm{\'a}rki and S. Bord{\'a}cs and P. Milde and E. Neuber and L. M. Eng and J. S. White and H. M. R{\o}nnow and C. D. Dewhurst and M. Mochizuki and K. Yanai and others},
  title     = {{N{\'e}el-Type Skyrmion Lattice with Confined Orientation in the Polar Magnetic Semiconductor GaV$_4$S$_8$}},
  journal   = {Nat. Mater.},
  volume    = {14},
  number    = {11},
  pages     = {1116--1122},
  year      = {2015},
  publisher = {Nature Publishing Group UK London},
}

@article{kurumaji2019skyrmion,
  author    = {T. Kurumaji and T. Nakajima and M. Hirschberger and A. Kikkawa and Y. Yamasaki and H. Sagayama and H. Nakao and Y. Taguchi and T.-H. Arima and Y. Tokura},
  title     = {{Skyrmion Lattice with a Giant Topological Hall Effect in a Frustrated Triangular-Lattice Magnet}},
  journal   = {Science},
  volume    = {365},
  number    = {6456},
  pages     = {914--918},
  year      = {2019},
  publisher = {American Association for the Advancement of Science},
}

@article{fujishiro2019topological,
  author    = {Y. Fujishiro and N. Kanazawa and T. Nakajima and X. Z. Yu and K. Ohishi and Y. Kawamura and K. Kakurai and T. Arima and H. Mitamura and A. Miyake and others},
  title     = {{Topological Transitions among Skyrmion- and Hedgehog-Lattice States in Cubic Chiral Magnets}},
  journal   = {Nat. Commun.},
  volume    = {10},
  number    = {1},
  pages     = {1059},
  year      = {2019},
  publisher = {Nature Publishing Group UK London},
}

@article{khanh2020nanometric,
  author    = {N. D. Khanh and T. Nakajima and X. Yu and S. Gao and K. Shibata and M. Hirschberger and Y. Yamasaki and H. Sagayama and H. Nakao and L. Peng and others},
  title     = {{Nanometric Square Skyrmion Lattice in a Centrosymmetric Tetragonal Magnet}},
  journal   = {Nat. Nanotechnol.},
  volume    = {15},
  number    = {6},
  pages     = {444--449},
  year      = {2020},
  publisher = {Nature Publishing Group UK London},
}

@article{zheng2023hopfion,
  author    = {F. Zheng and N. S. Kiselev and F. N. Rybakov and L. Yang and W. Shi and S. Bl{\"u}gel and R. E. Dunin-Borkowski},
  title     = {{Hopfion Rings in a Cubic Chiral Magnet}},
  journal   = {Nature},
  volume    = {623},
  number    = {7988},
  pages     = {718--723},
  year      = {2023},
  publisher = {Nature Publishing Group UK London},
}

@article{yu2018transformation,
  author    = {X. Z. Yu and W. Koshibae and Y. Tokunaga and K. Shibata and Y. Taguchi and N. Nagaosa and Y. Tokura},
  title     = {{Transformation between Meron and Skyrmion Topological Spin Textures in a Chiral Magnet}},
  journal   = {Nature},
  volume    = {564},
  number    = {7734},
  pages     = {95--98},
  year      = {2018},
  publisher = {Nature Publishing Group UK London},
}

@article{ohgushi2000spin,
  author    = {K. Ohgushi and S. Murakami and N. Nagaosa},
  title     = {{Spin Anisotropy and Quantum Hall Effect in the Kagom{\'e} Lattice: Chiral Spin State Based on a Ferromagnet}},
  journal   = {Phys. Rev. B},
  volume    = {62},
  number    = {10},
  pages     = {R6065},
  year      = {2000},
  publisher = {APS},
}

@article{taguchi2001spin,
  author    = {Y. Taguchi and Y. Oohara and H. Yoshizawa and N. Nagaosa and Y. Tokura},
  title     = {{Spin Chirality, Berry Phase, and Anomalous Hall Effect in a Frustrated Ferromagnet}},
  journal   = {Science},
  volume    = {291},
  number    = {5513},
  pages     = {2573--2576},
  year      = {2001},
  publisher = {American Association for the Advancement of Science},
}

@article{machida2010time,
  author    = {Y. Machida and S. Nakatsuji and S. Onoda and T. Tayama and T. Sakakibara},
  title     = {{Time-Reversal Symmetry Breaking and Spontaneous Hall Effect without Magnetic Dipole Order}},
  journal   = {Nature},
  volume    = {463},
  number    = {7278},
  pages     = {210--213},
  year      = {2010},
  publisher = {Nature Publishing Group UK London},
}

@article{tokura2021magnetic,
  author    = {Y. Tokura and N. Kanazawa},
  title     = {{Magnetic Skyrmion Materials}},
  journal   = {Chem. Rev.},
  volume    = {121},
  number    = {5},
  pages     = {2857--2897},
  year      = {2021},
  publisher = {ACS Publications},
}

@article{neubauer2009topological,
  author    = {A. Neubauer and C. Pfleiderer and B. Binz and A. Rosch and R. Ritz and P. G. Niklowitz and P. B{\"o}ni},
  title     = {{Topological Hall Effect in the A Phase of MnSi}},
  journal   = {Phys. Rev. Lett.},
  volume    = {102},
  number    = {18},
  pages     = {186602},
  year      = {2009},
  publisher = {APS},
}

@article{fujishiro2021giant,
  author    = {Y. Fujishiro and N. Kanazawa and R. Kurihara and H. Ishizuka and T. Hori and F. S. Yasin and X. Yu and A. Tsukazaki and M. Ichikawa and M. Kawasaki and others},
  title     = {{Giant Anomalous Hall Effect from Spin-Chirality Scattering in a Chiral Magnet}},
  journal   = {Nat. Commun.},
  volume    = {12},
  number    = {1},
  pages     = {317},
  year      = {2021},
  publisher = {Nature Publishing Group UK London},
}

@article{hirschberger2019skyrmion,
  author    = {M. Hirschberger and T. Nakajima and S. Gao and L. Peng and A. Kikkawa and T. Kurumaji and M. Kriener and Y. Yamasaki and H. Sagayama and H. Nakao and others},
  title     = {{Skyrmion Phase and Competing Magnetic Orders on a Breathing Kagom{\'e} Lattice}},
  journal   = {Nat. Commun.},
  volume    = {10},
  number    = {1},
  pages     = {5831},
  year      = {2019},
  publisher = {Nature Publishing Group UK London},
}

@article{takagi2023spontaneous,
  author    = {H. Takagi and R. Takagi and S. Minami and T. Nomoto and K. Ohishi and M.-T. Suzuki and Y. Yanagi and M. Hirayama and N. D. Khanh and K. Karube and others},
  title     = {{Spontaneous Topological Hall Effect Induced by Non-Coplanar Antiferromagnetic Order in Intercalated van der Waals Materials}},
  journal   = {Nat. Phys.},
  volume    = {19},
  number    = {7},
  pages     = {961--968},
  year      = {2023},
  publisher = {Nature Publishing Group UK London},
}

@article{park2023tetrahedral,
  author    = {P. Park and W. Cho and C. Kim and Y. An and Y.-G. Kang and M. Avdeev and R. Sibille and K. Iida and R. Kajimoto and K. H. Lee and others},
  title     = {{Tetrahedral Triple-$Q$ Magnetic Ordering and Large Spontaneous Hall Conductivity in the Metallic Triangular Antiferromagnet Co$_{1/3}$TaS$_2$}},
  journal   = {Nat. Commun.},
  volume    = {14},
  number    = {1},
  pages     = {8346},
  year      = {2023},
  publisher = {Nature Publishing Group UK London},
}

@article{martin2008itinerant,
  author    = {I. Martin and C. D. Batista},
  title     = {{Itinerant Electron-Driven Chiral Magnetic Ordering and Spontaneous Quantum Hall Effect in Triangular Lattice Models}},
  journal   = {Phys. Rev. Lett.},
  volume    = {101},
  number    = {15},
  pages     = {156402},
  year      = {2008},
  publisher = {APS},
}

@article{akagi2010spin,
  author    = {Y. Akagi and Y. Motome},
  title     = {{Spin Chirality Ordering and Anomalous Hall Effect in the Ferromagnetic Kondo Lattice Model on a Triangular Lattice}},
  journal   = {J. Phys. Soc. Jpn.},
  volume    = {79},
  number    = {8},
  pages     = {083711},
  year      = {2010},
  publisher = {The Physical Society of Japan},
}

@article{hayami2014multiple,
  author    = {S. Hayami and Y. Motome},
  title     = {{Multiple-$Q$ Instability by $(d-2)$-Dimensional Connections of Fermi Surfaces}},
  journal   = {Phys. Rev. B},
  volume    = {90},
  number    = {6},
  pages     = {060402},
  year      = {2014},
  publisher = {APS},
}

@article{wang2020skyrmion,
  author    = {Z. Wang and Y. Su and S.-Z. Lin and C. D. Batista},
  title     = {{Skyrmion Crystal from RKKY Interaction Mediated by 2D Electron Gas}},
  journal   = {Phys. Rev. Lett.},
  volume    = {124},
  number    = {20},
  pages     = {207201},
  year      = {2020},
  publisher = {APS},
}

@article{bouaziz2022fermi,
  author    = {J. Bouaziz and E. Mendive-Tapia and S. Bl{\"u}gel and J. B. Staunton},
  title     = {{Fermi-Surface Origin of Skyrmion Lattices in Centrosymmetric Rare-Earth Intermetallics}},
  journal   = {Phys. Rev. Lett.},
  volume    = {128},
  number    = {15},
  pages     = {157206},
  year      = {2022},
  publisher = {APS},
}

@article{paddison2022magnetic,
  author    = {J. A. M. Paddison and B. K. Rai and A. F. May and S. Calder and M. B. Stone and M. D. Frontzek and A. D. Christianson},
  title     = {{Magnetic Interactions of the Centrosymmetric Skyrmion Material Gd$_2$PdSi$_3$}},
  journal   = {Phys. Rev. Lett.},
  volume    = {129},
  number    = {13},
  pages     = {137202},
  year      = {2022},
  publisher = {APS},
}

@article{dong2025pseudogap,
  author    = {Y. Dong and Y. Kinoshita and M. Ochi and R. Nakachi and R. Higashinaka and S. Hayami and Y. Wan and Y. Arai and S. Huh and M. Hashimoto and others},
  title     = {{Pseudogap and Fermi Arc Induced by Fermi Surface Nesting in a Centrosymmetric Skyrmion Magnet}},
  journal   = {Science},
  volume    = {388},
  number    = {6747},
  pages     = {624--630},
  year      = {2025},
  publisher = {American Association for the Advancement of Science},
}

@article{arai2026origin,
  author    = {Y. Arai and K. Nakayama and A. Honma and S. Souma and D. Shiga and H. Kumigashira and T. Takahashi and K. Segawa and T. Sato},
  title     = {{Origin of Multiple Skyrmion Phases in EuAl$_4$}},
  journal   = {Nat. Commun.},
  volume    = {17},
  number    = {1},
  pages     = {3162},
  year      = {2026},
  publisher = {Nature Publishing Group UK London},
}

@article{jiang2015chiral,
  author    = {K. Jiang and Y. Zhang and S. Zhou and Z. Wang},
  title     = {{Chiral Spin Density Wave Order on the Frustrated Honeycomb and Bilayer Triangle Lattice Hubbard Model at Half-Filling}},
  journal   = {Phys. Rev. Lett.},
  volume    = {114},
  number    = {21},
  pages     = {216402},
  year      = {2015},
  publisher = {APS},
}

@article{ortiz2019new,
  author    = {B. R. Ortiz and L. C. Gomes and J. R. Morey and M. Winiarski and M. Bordelon and J. S. Mangum and I. W. H. Oswald and J. A. Rodriguez-Rivera and J. R. Neilson and S. D. Wilson and others},
  title     = {{New Kagome Prototype Materials: Discovery of KV$_3$Sb$_5$, RbV$_3$Sb$_5$, and CsV$_3$Sb$_5$}},
  journal   = {Phys. Rev. Mater.},
  volume    = {3},
  number    = {9},
  pages     = {094407},
  year      = {2019},
  publisher = {APS},
}

@article{Jiang_2021,
  author    = {Y.-X. Jiang and J.-X. Yin and M. M. Denner and N. Shumiya and B. R. Ortiz and G. Xu and Z. Guguchia and J. He and M. S. Hossain and X. Liu and J. Ruff and L. Kautzsch and S. S. Zhang and G. Chang and I. Belopolski and Q. Zhang and T. A. Cochran and D. Multer and M. Litskevich and Z.-J. Cheng and X. P. Yang and Z. Wang and R. Thomale and T. Neupert and S. D. Wilson and M. Z. Hasan},
  title     = {{Unconventional Chiral Charge Order in Kagome Superconductor KV$_3$Sb$_5$}},
  journal   = {Nat. Mater.},
  volume    = {20},
  number    = {10},
  pages     = {1353--1357},
  year      = {2021},
  month     = {Jun},
  issn      = {1476-4660},
  publisher = {Springer Science and Business Media LLC},
}

@article{Mielke_2022,
  author    = {C. Mielke and D. Das and J.-X. Yin and H. Liu and R. Gupta and Y.-X. Jiang and M. Medarde and X. Wu and H. C. Lei and J. Chang and P. Dai and Q. Si and H. Miao and R. Thomale and T. Neupert and Y. Shi and R. Khasanov and M. Z. Hasan and H. Luetkens and Z. Guguchia},
  title     = {{Time-Reversal Symmetry-Breaking Charge Order in a Kagome Superconductor}},
  journal   = {Nature},
  volume    = {602},
  number    = {7896},
  pages     = {245--250},
  year      = {2022},
  month     = {Feb},
  issn      = {1476-4687},
  publisher = {Springer Science and Business Media LLC},
}

@article{Kang_2022,
  author    = {M. Kang and S. Fang and J.-K. Kim and B. R. Ortiz and S. H. Ryu and J. Kim and J. Yoo and G. Sangiovanni and D. Di Sante and B.-G. Park and C. Jozwiak and A. Bostwick and E. Rotenberg and E. Kaxiras and S. D. Wilson and J.-H. Park and R. Comin},
  title     = {{Twofold van Hove Singularity and Origin of Charge Order in Topological Kagome Superconductor CsV$_3$Sb$_5$}},
  journal   = {Nat. Phys.},
  volume    = {18},
  number    = {3},
  pages     = {301--308},
  year      = {2022},
  month     = {Jan},
  issn      = {1745-2481},
  publisher = {Springer Science and Business Media LLC},
}

@article{Denner_2021,
  author    = {M. M. Denner and R. Thomale and T. Neupert},
  title     = {{Analysis of Charge Order in the Kagome Metal AV$_3$Sb$_5$ (A = K, Rb, Cs)}},
  journal   = {Phys. Rev. Lett.},
  volume    = {127},
  number    = {21},
  pages = {217601},
  year      = {2021},
  issn      = {1079-7114},
  publisher = {American Physical Society (APS)},
}

@article{Park_2021,
  author    = {T. Park and M. Ye and L. Balents},
  title     = {{Electronic Instabilities of Kagome Metals: Saddle Points and Landau Theory}},
  journal   = {Phys. Rev. B},
  volume    = {104},
  pages = {035142},
  number    = {3},
  year      = {2021},
  month     = {Jul},
  issn      = {2469-9969},
  publisher = {American Physical Society (APS)},
}

@article{Kohn_1967,
  author    = {W. Kohn},
  title     = {{Excitonic Phases}},
  journal   = {Phys. Rev. Lett.},
  volume    = {19},
  number    = {8},
  pages     = {439--442},
  year      = {1967},
  month     = {Aug},
  publisher = {American Physical Society (APS)},
}

@article{Jerome_1967,
  author    = {D. Jérome and T. M. Rice and W. Kohn},
  title     = {{Excitonic Insulator}},
  journal   = {Phys. Rev.},
  volume    = {158},
  number    = {2},
  pages     = {462--475},
  year      = {1967},
  month     = {Jun},
  publisher = {American Physical Society (APS)},
}

@article{HALPERIN_1968,
  author    = {B. I. Halperin and T. M. Rice},
  title     = {{Possible Anomalies at a Semimetal-Semiconductor Transition}},
  journal   = {Rev. Mod. Phys.},
  volume    = {40},
  number    = {4},
  pages     = {755--766},
  year      = {1968},
  month     = {Oct},
  publisher = {American Physical Society (APS)},
}

@article{Cercellier_2007,
  author    = {H. Cercellier and C. Monney and F. Clerc and C. Battaglia and L. Despont and M. G. Garnier and H. Beck and P. Aebi and L. Patthey and H. Berger and L. Forró},
  title     = {{Evidence for an Excitonic Insulator Phase in TiSe$_2$}},
  journal   = {Phys. Rev. Lett.},
  volume    = {99},
  number    = {14},
  pages = {146403},
  year      = {2007},
  month     = {Oct},
  issn      = {1079-7114},
  publisher = {American Physical Society (APS)},
}

@article{Monney_2009,
  author    = {C. Monney and H. Cercellier and F. Clerc and C. Battaglia and E. F. Schwier and C. Didiot and M. G. Garnier and H. Beck and P. Aebi and H. Berger and L. Forró and L. Patthey},
  title     = {{Spontaneous Exciton Condensation in 1T-TiSe$_2$: BCS-Like Approach}},
  journal   = {Phys. Rev. B},
  volume    = {79},
  number    = {4},
  pages = {045116},
  year      = {2009},
  month     = {Jan},
  publisher = {American Physical Society (APS)},
}

@article{Kogar_2017,
  author    = {A. Kogar and M. S. Rak and S. Vig and A. A. Husain and F. Flicker and Y. I. Joe and L. Venema and G. J. MacDougall and T. C. Chiang and E. Fradkin and J. van Wezel and P. Abbamonte},
  title     = {{Signatures of Exciton Condensation in a Transition Metal Dichalcogenide}},
  journal   = {Science},
  volume    = {358},
  number    = {6368},
  pages     = {1314--1317},
  year      = {2017},
  month     = {Dec},
  publisher = {American Association for the Advancement of Science (AAAS)},
}

@article{Wakisaka_2009,
  author    = {Y. Wakisaka and T. Sudayama and K. Takubo and T. Mizokawa and M. Arita and H. Namatame and M. Taniguchi and N. Katayama and M. Nohara and H. Takagi},
  title     = {{Excitonic Insulator State in Ta$_2$NiSe$_5$ Probed by Photoemission Spectroscopy}},
  journal   = {Phys. Rev. Lett.},
  volume    = {103},
  number    = {2},
  pages = {026402},
  year      = {2009},
  month     = {Jul},
  publisher = {American Physical Society (APS)},
}

@article{Lu_2017,
  author    = {Y. F. Lu and H. Kono and T. I. Larkin and A. W. Rost and T. Takayama and A. V. Boris and B. Keimer and H. Takagi},
  title     = {{Zero-Gap Semiconductor to Excitonic Insulator Transition in Ta$_2$NiSe$_5$}},
  journal   = {Nat. Commun.},
  volume    = {8},
  number    = {1},
  pages = {14408},
  year      = {2017},
  month     = {Feb},
  publisher = {Springer Science and Business Media LLC},
}

@article{Sun_2021,
  author    = {B. Sun and W. Zhao and T. Palomaki and Z. Fei and E. Runburg and P. Malinowski and X. Huang and J. Cenker and Y.-T. Cui and J.-H. Chu and X. Xu and S. S. Ataei and D. Varsano and M. Palummo and E. Molinari and M. Rontani and D. H. Cobden},
  title     = {{Evidence for Equilibrium Exciton Condensation in Monolayer WTe$_2$}},
  journal   = {Nat. Phys.},
  volume    = {18},
  number    = {1},
  pages     = {94--99},
  year      = {2021},
  month     = {Dec},
  publisher = {Springer Science and Business Media LLC},
}

@misc{Vylet_2026,
  author        = {K. Vylet and X. Huang and L. Balents},
  title         = {{Chiral Magnetism and Quantum Anomalous Hall Effect in a Low-Energy Kondo Model on the Triangular Lattice}},
  year          = {2026},
  eprint        = {2604.17641},
  archivePrefix = {arXiv},
  primaryClass  = {cond-mat.str-el},
  url           = {https://arxiv.org/abs/2604.17641},
}

@misc{Guzman_2026,
  author        = {K. Guzman and H. Ishizuka},
  title         = {{Field-Induced Metal-Insulator Transition, Chern Insulators, and Topological Semimetals in a Clean Magnetic Semiconductor GdGaI}},
  year          = {2026},
  eprint        = {2605.01804},
  archivePrefix = {arXiv},
  primaryClass  = {cond-mat.str-el},
  url           = {https://arxiv.org/abs/2605.01804},
}

@article{Okuma_2021,
  author    = {R. Okuma and C. Ritter and G. J. Nilsen and Y. Okada},
  title     = {{Magnetic Frustration in a van der Waals Metal CeSiI}},
  journal   = {Phys. Rev. Mater.},
  volume    = {5},
  number    = {12},
  pages = {L121401},
  year      = {2021},
  month     = {Dec},
  publisher = {American Physical Society (APS)},
}

@article{Posey_2024,
  author    = {V. A. Posey and S. Turkel and M. Rezaee and A. Devarakonda and A. K. Kundu and C. S. Ong and M. Thinel and D. G. Chica and R. A. Vitalone and R. Jing and S. Xu and D. R. Needell and E. Meirzadeh and M. L. Feuer and A. Jindal and X. Cui and T. Valla and P. Thunström and T. Yilmaz and E. Vescovo and D. Graf and X. Zhu and A. Scheie and A. F. May and O. Eriksson and D. N. Basov and C. R. Dean and A. Rubio and P. Kim and M. E. Ziebel and A. J. Millis and A. N. Pasupathy and X. Roy},
  title     = {{Two-Dimensional Heavy Fermions in the van der Waals Metal CeSiI}},
  journal   = {Nature},
  volume    = {625},
  number    = {7995},
  pages     = {483--488},
  year      = {2024},
  month     = {Jan},
  publisher = {Springer Science and Business Media LLC},
}

@article{Lukachuk_2007,
  author    = {M. Lukachuk and C. Zheng and H. Mattausch and R. K. Kremer and A. Simon and M. G. Banks},
  title     = {{RE$_{2+x}$I$_2$M$_{2+y}$ (RE = Ce, Gd, Y; M = Al, Ga): Reduced Rare Earth Halides with a Hexagonal Metal Atom Network}},
  journal   = {Z. Naturforsch. B},
  volume    = {62},
  number    = {5},
  pages     = {633--641},
  year      = {2007},
  month     = {May},
  publisher = {Walter de Gruyter GmbH},
}

@article{Kaneko_2026,
  author    = {T. Kaneko and R. Mizuno and S. Kamiyama and H. Miyamoto and M. Ochi},
  title     = {{Electronic Band Structure, Phonon Dispersion, and Magnetic Triple-$Q$ State in GdGaI}},
  journal   = {Phys. Rev. B},
  volume    = {113},
  number    = {4},
  pages = {045156},
  year      = {2026},
  month     = {Jan},
  publisher = {American Physical Society (APS)},
}

@misc{okuma2024,
      title={Emergent topological magnetism in Hund's excitonic insulator}, 
      author={R. Okuma and K. Yamagami and Y. Fujisawa and C. H. Hsu and Y. Obata and N. Tomoda and M. Dronova and K. Kuroda and H. Ishikawa and K. Kawaguchi and K. Aido and K. Kindo and Y. H. Chan and H. Lin and Y. Ihara and T. Kondo and Y. Okada},
      year={2024},
      eprint={2405.16781},
      archivePrefix={arXiv},
      primaryClass={cond-mat.str-el},
      url={https://arxiv.org/abs/2405.16781}, 
}

@article{Akagi_2012,
  author    = {Y. Akagi and M. Udagawa and Y. Motome},
  title     = {{Hidden Multiple-Spin Interactions as an Origin of Spin Scalar Chiral Order in Frustrated Kondo Lattice Models}},
  journal   = {Phys. Rev. Lett.},
  volume    = {108},
  number    = {9},
  pages = {096401},
  year      = {2012},
  month     = {Feb},
  publisher = {American Physical Society (APS)},
}

@article{Nakatsuji_2015,
  author    = {S. Nakatsuji and N. Kiyohara and T. Higo},
  title     = {{Large Anomalous Hall Effect in a Non-Collinear Antiferromagnet at Room Temperature}},
  journal   = {Nature},
  volume    = {527},
  number    = {7577},
  pages     = {212--215},
  year      = {2015},
  month     = {Oct},
  doi       = {10.1038/nature15723},
  url       = {https://doi.org/10.1038/nature15723},
  issn      = {1476-4687},
  publisher = {Springer Science and Business Media LLC},
}

@article{Nayak_2016,
  author    = {A. K. Nayak and J. E. Fischer and Y. Sun and B. Yan and J. Karel and A. C. Komarek and C. Shekhar and N. Kumar and W. Schnelle and J. Kübler and C. Felser and S. S. P. Parkin},
  title     = {{Large Anomalous Hall Effect Driven by a Nonvanishing Berry Curvature in the Noncolinear Antiferromagnet Mn$_3$Ge}},
  journal   = {Sci. Adv.},
  volume    = {2},
  number    = {4},
  pages = {e1501870},
  year      = {2016},
  month     = {Apr},
  publisher = {American Association for the Advancement of Science (AAAS)},
}

@article{Kiyohara_2016,
  author    = {N. Kiyohara and T. Tomita and S. Nakatsuji},
  title     = {{Giant Anomalous Hall Effect in the Chiral Antiferromagnet Mn$_3$Ge}},
  journal   = {Phys. Rev. Appl.},
  volume    = {5},
  number    = {6},
  pages     = {064009},
  year      = {2016},
  month     = {Jun},
  publisher = {American Physical Society (APS)},
}

@article{LeeBroken2024,
  author    = {S. Lee and S. Lee and S. Jung and J. Jung and D. Kim and Y. Lee and B. Seok and J. Kim and B. G. Park and L. {\v{S}}mejkal and C.-J. Kang and C. Kim},
  title     = {{Broken Kramers Degeneracy in Altermagnetic MnTe}},
  journal   = {Phys. Rev. Lett.},
  volume    = {132},
  number    = {3},
  pages     = {036702},
  year      = {2024},
  month     = {Jan},
  numpages  = {7},
  publisher = {American Physical Society},
}

@article{TakagiFeS_2024,
  author    = {R. Takagi and R. Hirakida and Y. Settai and R. Oiwa and H. Takagi and A. Kitaori and K. Yamauchi and H. Inoue and J.-I. Yamaura and D. Nishio-Hamane and S. Itoh and S. Aji and H. Saito and T. Nakajima and T. Nomoto and R. Arita and S. Seki},
  title     = {{Spontaneous Hall Effect Induced by Collinear Antiferromagnetic Order at Room Temperature}},
  journal   = {Nat. Mater.},
  volume    = {24},
  number    = {1},
  pages     = {63--68},
  year      = {2024},
  month     = {Dec},
  publisher = {Springer Science and Business Media LLC},
}

@article{Yamadapwave_2025,
  author    = {R. Yamada and M. T. Birch and P. R. Baral and S. Okumura and R. Nakano and S. Gao and M. Ezawa and T. Nomoto and J. Masell and Y. Ishihara and K. K. Kolincio and I. Belopolski and H. Sagayama and H. Nakao and K. Ohishi and T. Ohhara and R. Kiyanagi and T. Nakajima and Y. Tokura and T.-H. Arima and Y. Motome and M. M. Hirschmann and M. Hirschberger},
  title     = {{A Metallic $p$-Wave Magnet with Commensurate Spin Helix}},
  journal   = {Nature},
  volume    = {646},
  number    = {8086},
  pages     = {837--842},
  year      = {2025},
  month     = {Oct},
  issn      = {1476-4687},
  publisher = {Springer Science and Business Media LLC},
}

@article{Jiang_2026,
  author    = {N. Jiang and Y. Zhang and Y. Xia and T. Higashihara and R. Okuma and J.-I. Yamaura and Y. Okada and K. Watanabe and T. Taniguchi and T. Zhang and T. Machida and Y. Niimi},
  title     = {{Raman Spectroscopy of the van der Waals Topological Magnet GdGaI}},
  journal   = {Phys. Rev. B},
  volume    = {114},
  number    = {6},
  pages = {065128},
  year      = {2026},
  month     = {Jul},
  issn      = {2469-9969},
  publisher = {American Physical Society (APS)},
}

@article{HayamiCDW2021,
  author    = {S. Hayami and Y. Motome},
  title     = {{Charge Density Waves in Multiple-$Q$ Spin States}},
  journal   = {Phys. Rev. B},
  volume    = {104},
  number    = {14},
  pages = {144404},
  year      = {2021},
  month     = {Oct},
  publisher = {American Physical Society (APS)},
}

@article{Barros_2014,
  author    = {K. Barros and J. W. F. Venderbos and G.-W. Chern and C. D. Batista},
  title     = {{Exotic Magnetic Orderings in the Kagome Kondo-Lattice Model}},
  journal   = {Phys. Rev. B},
  volume    = {90},
  number    = {24},
  pages = {245119},
  year      = {2014},
  month     = {Dec},
  publisher = {American Physical Society (APS)},
}

@article{Miyasato_2007,
  author    = {T. Miyasato and N. Abe and T. Fujii and A. Asamitsu and S. Onoda and Y. Onose and N. Nagaosa and Y. Tokura},
  title     = {{Crossover Behavior of the Anomalous Hall Effect and Anomalous Nernst Effect in Itinerant Ferromagnets}},
  journal   = {Phys. Rev. Lett.},
  volume    = {99},
  number    = {8},
  pages = {086602},
  year      = {2007},
  month     = {Aug},
  publisher = {American Physical Society (APS)},
}

@article{Tian_2009,
  author    = {Y. Tian and L. Ye and X. Jin},
  title     = {{Proper Scaling of the Anomalous Hall Effect}},
  journal   = {Phys. Rev. Lett.},
  volume    = {103},
  number    = {8},
  pages = {087206},
  year      = {2009},
  month     = {Aug},
  publisher = {American Physical Society (APS)},
}

@article{OnodaAHE2006,
  author    = {S. Onoda and N. Sugimoto and N. Nagaosa},
  title     = {{Intrinsic versus Extrinsic Anomalous Hall Effect in Ferromagnets}},
  journal   = {Phys. Rev. Lett.},
  volume    = {97},
  number    = {12},
  pages     = {126602},
  year      = {2006},
  month     = {Sep},
  numpages  = {4},
  publisher = {American Physical Society},
}

@article{Huckel_1931,
  author    = {E. Huckel},
  title     = {{Quantentheoretische Beitrage zum Benzolproblem: I. Die Elektronenkonfiguration des Benzols und Verwandter Verbindungen}},
  journal   = {Z. Phys.},
  volume    = {70},
  number    = {3-4},
  pages     = {204--286},
  year      = {1931},
  month     = {Mar},
  publisher = {Springer Science and Business Media LLC},
}

@article{Pauling_1934,
  author  = {L. Pauling and G. W. Wheland},
  title   = {{The Nature of the Chemical Bond. V.}},
  journal = {J. Chem. Phys.},
  volume  = {2},
  number  = {8},
  pages   = {482--482},
  year    = {1934},
  month   = {Aug},
}

@article{Mo_2009,
  author  = {Y. Mo},
  title   = {{The Resonance Energy of Benzene: A Revisit}},
  journal = {J. Phys. Chem. A},
  volume  = {113},
  number  = {17},
  pages   = {5163--5169},
  year    = {2009},
  month   = {Mar},
}

@article{momma2011vesta,
  author    = {K. Momma and F. Izumi},
  title     = {{VESTA 3 for Three-Dimensional Visualization of Crystal, Volumetric and Morphology Data}},
  journal   = {J. Appl. Cryst.},
  volume    = {44},
  number    = {6},
  pages     = {1272--1276},
  year      = {2011},
  publisher = {International Union of Crystallography},
}

@article{RASTELLI19791,
title = {Non-simple magnetic order for simple Hamiltonians},
journal = {Physica B+C},
volume = {97},
number = {1},
pages = {1-24},
year = {1979},
issn = {0378-4363},
author = {E. Rastelli and A. Tassi and L. Reatto},
}

@article{jolicoeur1990ground,
  author    = {T. Jolicoeur and E. Dagotto and E. Gagliano and S. Bacci},
  title     = {{Ground-State Properties of the $S=1/2$ Heisenberg Antiferromagnet on a Triangular Lattice}},
  journal   = {Phys. Rev. B},
  volume    = {42},
  number    = {7},
  pages     = {4800},
  year      = {1990},
  publisher = {APS},
}

@article{chubukov1992order,
  author    = {A. V. Chubukov and T. Jolicoeur},
  title     = {{Order-from-Disorder Phenomena in Heisenberg Antiferromagnets on a Triangular Lattice}},
  journal   = {Phys. Rev. B},
  volume    = {46},
  number    = {17},
  pages     = {11137},
  year      = {1992},
  publisher = {APS},
}
\end{document}